\documentclass{appolb}
\usepackage{epsfig}
\usepackage{cite}
\begin{document}
\pagestyle{plain}
\title{\bf THE METHOD TO DETERMINE \\ THE ${\cal CP}$ NATURE OF HIGGS BOSONS \\
FROM  DECAYS TO TAU LEPTONS  AT LC
\thanks{Presented at the  XXVII International Conference 
of Theoretical Physics,
{\it ``Matter To The Deepest''}, Ustro\'n, Poland, 15-21 September 2003. } 
%~\thanks{{\tt hep-ph/0305082}, {\tt LC-PHSM-2003-050}, {\tt TTP03-13}}
}
\author{Ma\l gorzata Worek 
\address{
Institute of Nuclear Physics of the  Polish Academy of Sciences
\\ Radzikowskiego 152,
31-342 Krakow, Poland.
\\ and
\\ Institute of  Physics, University of Silesia \\ Uniwersytecka 4,
 40-007 Katowice, Poland. \\ e-mail: {\tt Malgorzata.Worek@ifj.edu.pl}}
}
\maketitle

\begin{abstract}
We demonstrate how the transverse $\tau^{+}\tau^{-}$ spin correlations  
can be used  to 
determine whether a decaying Higgs boson is a mixed ${\cal CP}$
eigenstate, thereby directly probe the presence of 
${\cal CP}$ violation in the neutral Higgs boson sector.
We  investigate the subsequent decay chain
$H \to
\tau^{+}\tau^{-} \to \rho^{+}\bar{\nu}_{\tau}\rho^{-}\nu_{\tau} \to
\pi^{+}\pi^0\bar{\nu}_{\tau}\pi^{-}\pi^0\nu_{\tau}$.  The prospects for the
measurement of the pseudoscalar admixture in the $H\tau^{+}\tau^{-}$
coupling to a Standard Model Higgs boson with a mass  of $120$ $GeV$ are
quantified for the case of $e^{+}e^{-}$ collisions at  $350$ $GeV$
center-of-mass energy and $1$ $ab^{-1}$ integrated luminosity.
The Standard Model Higgsstrahlung production
process is used as an example. 
 
\end{abstract}
\section{Introduction}

A most distinctive feature of the extended 
models of Standard Model (${\cal SM}$) 
such as Minimal Supersymmetric Standard Model (${\cal MSSM}$), 
or general Two Higgs Boson
Doublet Model ($2{\cal HDM}$), 
is the existence of additional Higgs bosons, \ie
a charged Higgs boson pair, $H^{\pm}$, one  $\mathcal{CP}$ odd scalar
$A^{0}$ and two $\mathcal{CP}$ even scalars $h^{0},H^{0}$.
Their mass and coupling patterns vary with the model parameters. 
The discovery of a neutral Higgs boson  with mass in the range 
$114$ $GeV < M_{H}< 140$ $GeV$  will raise the question whether 
the observed particle is the ${\cal SM}$ Higgs boson or the lightest 
boson from the Higgs boson sector of a ${\cal SM}$ extension.
Even if the one doublet ${\cal SM}$ turns out to be a good model, we will
want to strictly determine that the observed Higgs 
boson is indeed ${\cal CP}$ even in 
its nature. 

Whichever scenario is realized in Nature, the major goal of future
high energy experiments will be  to distinguish   
among these models. The precision measurements of all relevant Higgs boson 
couplings are powerful to obtain information about additional Higgs 
doublets, their structure and masses. 
Such measurements can only be performed 
at the $e^{+}e^{-}$ linear collider ({\it LC}) such as {\it TESLA}. 
The determination 
of the quantum numbers of the Higgs boson(s) and sensitivity to the 
${\cal CP}$ violation represent a crucial part of this test and
will allow to establish 
the Higgs boson mechanism as the mechanism of the
 electroweak symmetry breaking. 

The determination of the parity and  the parity mixing of spinless Higgs
 bosons have been extensively investigated in 
Refs.~\cite{Dell'Aquila:1986ve,Nelson:1988ki,Kramer:1994jn,Barger:1994wt,Grzadkowski:1995rx,Hagiwara2000,Miller:2001bi,Dova:2001sq}. The model independent 
identification of parity of the Higgs particle has recently been demonstrated 
for $H/A^{0}\to\tau^{+}\tau^{+}$ decay chain in 
Refs.~\cite{Was:2002gv,Bower:2002zx,Desch:2003mw,Worek:2003zp,Desch:2003rw}.

In Ref. \cite{Was:2002gv}, see also 
Ref. \cite{Worek:2003zp} the reaction chain   $e^+e^-
\to Z (H/A^{0})$,     $H/A^{0} \to \tau^{+}\tau^{-}$,   $\tau^{\pm}\to
\pi^{\pm}\bar{\nu}_{\tau}({\nu}_{\tau})$ was    studied. It was found
that even small effects of smearing   seriously deteriorate the
measurement resolution.  However, the 
spin effects of the decay chain $H/A^{0} \to
\tau^{+}\tau^{-} \to \rho^{+}\bar{\nu}_{\tau}\rho^{-}\nu_{\tau} \to
\pi^{+}\pi^0\bar{\nu}_{\tau}\pi^{-}\pi^0\nu_{\tau}$  give a parity test
independent of both model (\eg ${\cal SM}$, ${\cal MSSM}$) and Higgs boson
production  mechanism (\eg Higgsstrahlung, WW fusion).  Using
reasonable assumptions about the ${\cal SM}$ production cross section and about
the measurement resolutions we have found that,   with $500$ $fb^{-1}$
of luminosity at a  $500$ $GeV$ $e^+e^-$ linear collider, the $\mathcal
{CP}$ of a $120$  $GeV$ Higgs boson can be measured to a confidence level
greater     than $95\%$, see Ref.~\cite{Bower:2002zx}.  

In Ref.~\cite{Desch:2003mw} we  demonstrated that a measurement of the
$\tau$ impact parameter  in one-prong $\tau$ decay is useful for the
determination of the  Higgs boson parity in the same decay chain. 
For a detection set-up such as  {\it TESLA}, use of the
information from the $\tau$ impact parameter can improve the
significance of the measurement of the parity of a Standard Model
$120$ $GeV$ Higgs boson to $\sim$ $4.5\sigma$.

In this paper we  continue to investigate the case of a Higgs boson 
decay into $\tau^{+}\tau^{-}$ and we 
 concentrate  on the more general case where mixed scalar and
pseudoscalar couplings of the  Higgs boson to $\tau$ leptons are
simultaneously allowed, see Ref.~\cite{Desch:2003rw} for details study.

The rest of the paper is
organized as follows.
In Section \ref{section_1} we present general information about measurement  
of the ${\cal CP}$ quantum numbers of Higgs boson using $\tau\tau$
decay channel. In section \ref{section_2} we recall basic properties of the 
density matrix for the pair of $\tau$ leptons produced in Higgs boson decay. 
In Section  \ref{section_3} we define an observable. Numerical results are 
presented in Section \ref{section_4}. The Summary 
closes the paper.
 
%%%%%%%%%%%%%%%%%%%%%%%%%%%%%%%%%%%%%%%%%%%%%%%%%%%%%%%%%%%%%%%%%%%
%                     new  section                                %
%%%%%%%%%%%%%%%%%%%%%%%%%%%%%%%%%%%%%%%%%%%%%%%%%%%%%%%%%%%%%%%%%%%
\section{ ${\cal CP}$ properties of the Higgs boson
 from {\boldmath $\tau\tau$} decay}
\label{section_1}

Various more or  less
complicated extensions of the ${\cal SM}$ have been proposed in 
literature, see \eg 
Ref.~\cite{2HDM,Weinberg:1990me,Pilaftsis:1998dd,Demir:1999hj,Ginzburg:2001ph,Ginzburg:2002wt,Niezurawski:2003ik}, 
but their common feature is that in case of models with 
${\cal CP}$ violation  in the Higgs boson sector
the mass eigenstates of the neutral Higgs bosons 
are not precisely ${\cal CP}$ eigenstates $h^{0}$, $H^{0}$ and $A^{0}$.
${\cal CP}$ violation results in three neutral states of 
mixed ${\cal CP}$ character.
For these models to estimate precision of the $H-A^{0}$ mixing angle 
is essential.  
We consider a potential of establishing ${\cal CP}$ properties of Higgs 
bosons,
without any assumption  about  a ${\cal CP}$ violation model, 
from the analysis of the angular distributions of the $\tau^{+}\tau^{-}$
decay products in the plane transverse to the    $\tau^{+}\tau^{-}$
axes. The transverse spin effects 
in $\tau$ pair production, reflected in correlations between $\tau$ 
decay products,  are  helpful to   distinguish between the scalar
$\mathcal{J^{PC}}=0^{++}$, pseudoscalar $\mathcal{J^{PC}}=0^{-+}$ and mixing
natures of the spin zero (Higgs) particles. 
When the Higgs boson  is light
enough that the $W^+W^-$ decay channel remains closed, 
the most promising decay channel of ${\cal SM}$ neutral Higgs 
particle sensitive for spin correlations, 
is  the $\tau^+\tau^-$ mode. 
The $\tau^+\tau^-$ channel
is useful in the ${\cal SM}$ for Higgs masses less than $\sim 140$
$GeV$. Up to this mass, the Higgs particle is very narrow,
$\Gamma(H_{SM})\le 10$ $MeV$ with 
 ${\cal BR}(H_{SM}\to \tau^{+}\tau^{-})\sim$ $9\%$.
In Supersymmetric theories, the  $\tau^+\tau^-$
channel is useful over a much larger mass range.
In this approach the ${\cal CP}$ properties of Higgs boson 
can be studied independently of a production mechanism and the specific
model and can be considered as the most general one.

It is generally believed that a Monte Carlo simulation  of
the full chain from the beam collision to detector response is the
most  convenient technique to investigate such studies.
In our analysis, we will 
take as an example the $e^{+}e^{-}\to ZH$; 
$Z\to \mu^{+} \mu^{-}$; $H\to \tau^{+} \tau^{-}$   
production process. 
We discuss a method for the ${\cal CP}$ quantum numbers
measurement of
the Higgs boson  with a mass  of $120$ $GeV$ for the case 
of $e^{+}e^{-}$ collisions with  $\sqrt{s}=350$ $GeV$
using  $H/A^0\to\tau^+\tau^-$;
$\tau^\pm\to\rho^\pm\bar{\nu}_{\tau}(\nu_{\tau})$;
$\rho^\pm\to\pi^\pm\pi^0$ decay chain.  
All the Monte Carlo samples have been
generated  with the {\tt TAUOLA} Monte Carlo 
library \cite{Jadach:1990mz,Jezabek:1991qp,Jadach:1993hs}.  
For the production of the $\tau$ lepton pairs the Monte Carlo program   
{\tt PYTHIA 6.1}  is used \cite{Sjostrand:2000wi}. 
The effects of initial state   
bremsstrahlung were included in the {\tt PYTHIA} generation.  
For  the $\tau$ lepton pair decay   
with full spin effects included in the $H\to\tau^{+}\tau^{-}$; 
$\tau^{\pm}\to\rho^{\pm}\bar{\nu}_{\tau}(\nu_{\tau})$; 
$\rho^{\pm}\to\pi^{\pm}\pi^{0}$ chain, the  interface explained in  
Refs.\cite{Was:2002gv,Bower:2002zx} was used. It is an extended version of 
the standard universal interface presented in
Ref.\cite{Pierzchala:2001gc}, see also Ref.\cite{Worek:2001hn}. 

%%%%%%%%%%%%%%%%%%%%%%%%%%%%%%%%%%%%%%%%%%%%%%%%%%%%%%%%%%%%%%%%%%%
%                     new  section                                %
%%%%%%%%%%%%%%%%%%%%%%%%%%%%%%%%%%%%%%%%%%%%%%%%%%%%%%%%%%%%%%%%%%%
\section{Mixed scalar--pseudoscalar case in Monte Carlo algorithm}
\label{section_2}

Let us now,  only very briefly repeat the basic information about the
spin correlations and their implementation in our Monte Carlo
algorithm. The detailed description of the method can be found in 
Ref.~\cite{Jadach:1990mz} and the full description of the algorithm is given 
in Ref.~\cite{Was:2002gv}.
 
The main  spin weight of our algorithm for generating the physical
process of $\tau$ lepton  pair production in Higgs boson decay, with
subsequent decay of $\tau$ leptons as well,  is given by
\begin{equation}
wt=\frac{1}{4}\left(1+\sum^{3}_{i=1}\sum^{3}_{j=1} R_{ij}h_{1}^{i}
h_{2}^{j}\right),    
\end{equation}
where $h_{1}$ and $h_{2}$  are the polarimeter vectors  that depend
respectively on $\tau^{\pm}$ decay products momenta; $R_{ij}$ is the
spin density  matrix. For the mixed scalar--pseudoscalar case, when the
general Higgs  boson Yukawa coupling to the $\tau$ lepton
\begin{equation}
\label{coupl}
\bar{\tau}(a+ib\gamma_{5})\tau
\end{equation}  
is assumed, we get the following non-zero components of $R_{ij}$, 
see Ref.~\cite{Desch:2003rw}:
\begin{equation}
R_{33}=-1,~~~~~
R_{11}=R_{22}=\frac{a^{2}\beta^{2}-b^{2}}{a^{2}\beta^{2}+b^{2}},~~~~~~~
R_{12}=-R_{21}=\frac{2ab\beta}{a^{2}\beta^{2}+b^{2}},
\end{equation}
where $\beta=\sqrt{1-{4m^{2}_{\tau}}/{m^{2}_{H}}}$. 
The crucial point is that, in general, $a$ and $b$ are of comparable 
magnitude in a ${\cal CP}$ violating extension of ${\cal SM}$.
For a ${\cal CP}$ conserving Higgs sector, either $a=0$ or $b=0$ implying 
a  pure
pseudoscalar or  scalar case respectively. For a ${\cal CP}$ mixed 
eigenstate, both $a$ and $b$ are non zero.  Thus any significant deviation of 
$R_{12}$ or equivalently $R_{21}$ from zero  
provides a signature for ${\cal CP}$ violation in the Higgs sector 
independent of the specific model.
 If we express
Eq.~(\ref{coupl}) with the help of the scalar--pseudoscalar mixing
angle $\phi$:
\begin{equation}
\label{coupla}
\bar{\tau}N(\cos\phi+i\sin\phi\gamma_{5})\tau,
\end{equation}
the  components of the spin density matrix can be expressed in the
following way:    
\begin{equation}
R_{11}=R_{22}=\frac{\cos\phi^{2}~\beta^{2}-
\sin\phi^{2}}{\cos\phi^{2}~\beta^{2}+\sin\phi^{2}}, ~~~~~~~
R_{12}=-R_{21}=\frac{2\cos\phi \sin\phi~\beta}
{\cos\phi^{2}~\beta^{2}+\sin\phi^{2}}.
\end{equation}
In  the limit $\beta \to 1$ these expressions reduce to the components
of the rotation  matrix for the rotation around the $z$ axis by an
angle $-2\phi$:
\begin{equation}
R_{11}=R_{22}=\cos2\phi, ~~~~~~~ R_{12}=-R_{21}=\sin2\phi.
\end{equation}

%%%%%%%%%%%%%%%%%%%%%%%%%%%%%%%%%%%%%%%%%%%%%%%%%%%%%%%%%%%%%%%%%%%
%                     new  section                                %
%%%%%%%%%%%%%%%%%%%%%%%%%%%%%%%%%%%%%%%%%%%%%%%%%%%%%%%%%%%%%%%%%%%
\section{Definition of the observable} 
\label{section_3}

Let us now recall a observable which we have introduced
to distinguish between scalar--pseudoscalar mixed state of the Higgs boson.
The method relies on measuring the acoplanarity angle
of the two planes, spanned on  $\rho^{\pm}$ decay products and defined
in the $\rho^{+}\rho^{-}$ pair rest frame.  For that purpose the
four-momenta of $\pi^{\pm}$ and $\pi^{0}$ need to be reconstructed
and, combined, they will  yield the $\rho^{\pm}$ four-momenta.  All
reconstructed four-momenta are then boosted into the
$\rho^{+}\rho^{-}$ pair rest frame. The acoplanarity angle
$\varphi^{*}$, between the planes of the $\rho^{+}$ and $\rho^{-}$
decay products is defined in this frame.
The correlation, in the case of the Higgs boson
of combined scalar and pseudoscalar couplings with
 the mixing angle $\phi$, is between transverse components of
$\tau^+$ spin polarization vector and transverse components of
$\tau^-$ polarization vector rotated by an angle
$2\phi$. Therefore  the full range of the variable $0 <\varphi^{*} <
2\pi$ is of  physical  interest. To distinguish between the two cases
$\varphi^{*}$ and  $2\pi-\varphi^{*}$ it is sufficient, for example,
to find the sign of    
$p_{\pi^-} \cdot {\bf n}_+$, where ${\bf n}_+$ is a vector normal to 
the plane spanned by the visible decay products of $\rho^{+}$, 
${\bf n}_+={\bf p}_{\pi^{+}}\times{\bf p}_{\pi^{0}}$. 
The range $0 <\varphi^{*} < \pi$ corresponds to the
negative sign case, otherwise one should make the
replacement $\varphi^{*} \rightarrow 2\pi-\varphi^{*}$.
If no separation was made, the parity effect, in case of mixed
$H\tau\tau$ coupling, would wash itself out.  
Additional selection cuts  need to be applied. Otherwise the acoplanarity
distribution is not sensitive to transverse spin effects at all.  
The events need to be  divided into two
classes, depending on the sign of  $y_{1}y_{2}$, where
\begin{equation}
y_1={E_{\pi^{+}}-E_{\pi^{0}}\over E_{\pi^{+}}+E_{\pi^{0}}}~;~~~~~
y_2={E_{\pi^{-}}-E_{\pi^{0}}\over E_{\pi^{-}}+E_{\pi^{0}}}.
\label{y1y2}
\end{equation}
The energies of $\pi^\pm,\pi^0$ are to be taken in the  respective
$\tau^\pm$ rest frames. In Refs.~\cite{Bower:2002zx,Desch:2003mw} the
methods of reconstruction of the replacement $\tau^\pm$ rest frames were
proposed with and without the help of the $\tau$ impact parameter.  We
will use these methods here as well, without any modification.

%%%%%%%%%%%%%%%%%%%%%%%%%%%%%%%%%%%%%%%%%%%%%%%%%%%%%%%%%%%%%%%%%%%
%                     new  section                                %
%%%%%%%%%%%%%%%%%%%%%%%%%%%%%%%%%%%%%%%%%%%%%%%%%%%%%%%%%%%%%%%%%%%
\section{Numerical results} 
\label{section_4}

%*******************************************************************  
\begin{figure}[!ht]
\begin{center} 
\epsfig{file=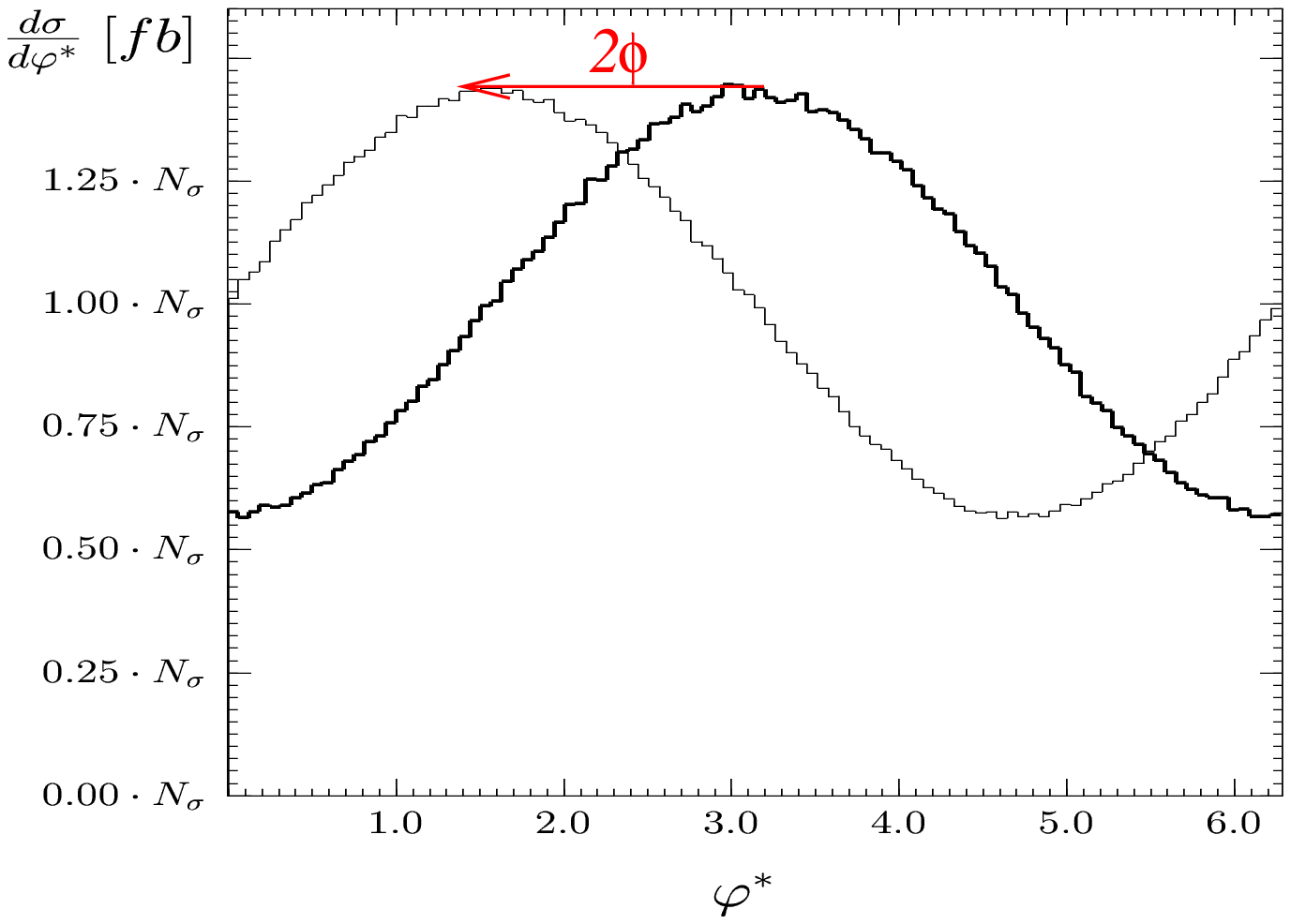,width=60mm,height=60mm}
\epsfig{file=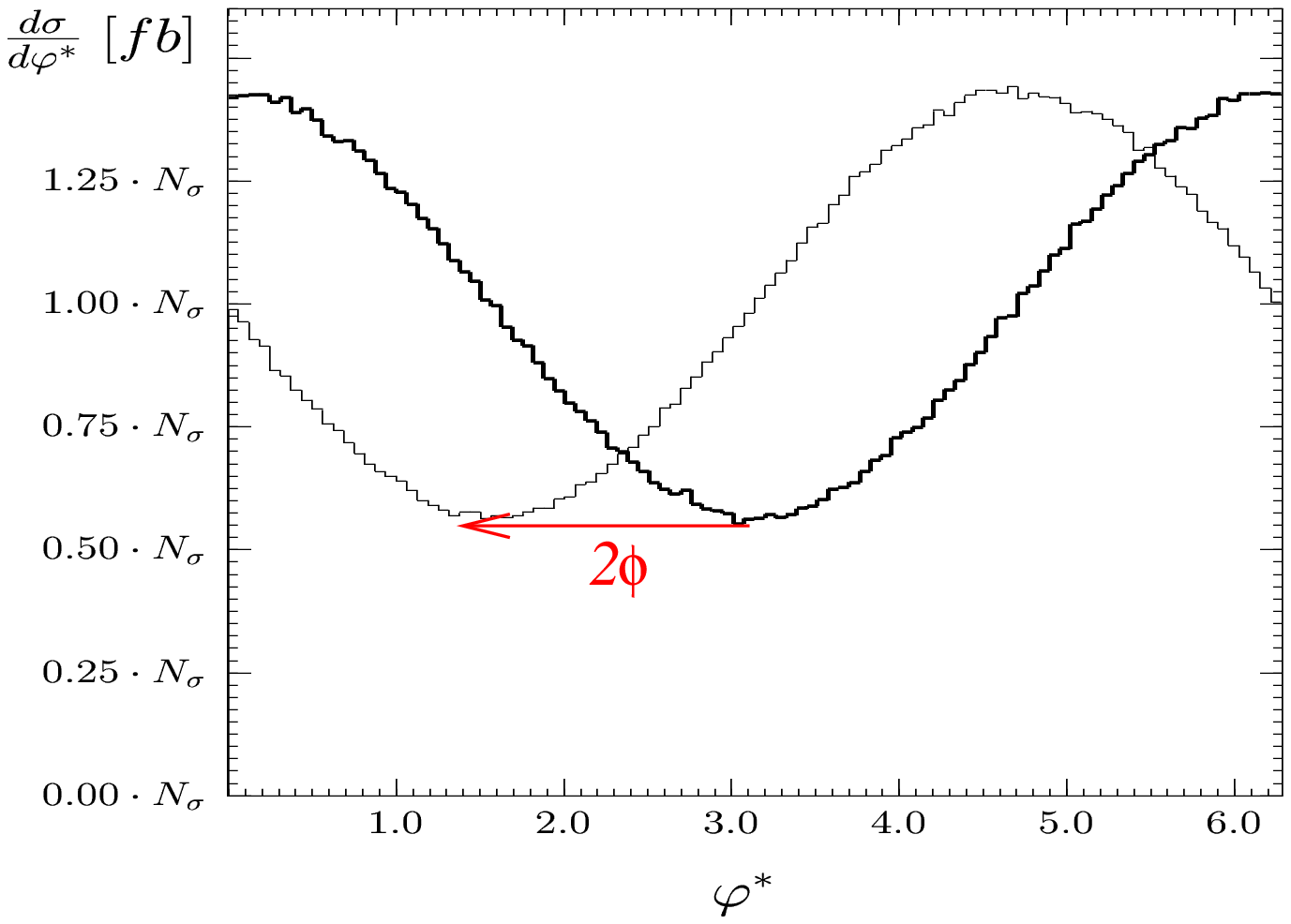,width=60mm,height=60mm}
\end{center} 
\vspace{-0.5 cm}
\caption  
{\it  The acoplanarity distribution  (angle $\varphi^{*}$) of the
$\rho^+ \rho^-$ decay products     in the rest frame of the $\rho^+
\rho^-$ pair. Gaussian smearing of $\pi$'s momenta  and generator level
$\tau^\pm$ rest frames are used.  The thick line corresponds to a
scalar Higgs boson,  the thin line to a mixed one.   The left figure
contains events with $y_1 y_2 > 0$, the right one is for $y_1 y_2 < 0$.
}
\label{rys1}  
\end{figure}  
%*******************************************************************

Let us
turn our attention to the measurable distributions.
We have used the scalar--pseudoscalar mixing angle
$\phi=\frac{\pi}{4}$ and,  as the reference, we have used the  pure
scalar case $\phi=0$. In our study, 
that is for the  $350$ $GeV$ $e^+e^-$  
center-of-mass energy, Higgs boson mass of
$120$ $GeV$ and 
 Higgsstrahlung production process, we took
$N_\sigma=62.7 \cdot 10^{-3}\; [fb]$ for the scale of the plot. 
However, in general case
\begin{equation} 
N_\sigma=\frac{1}{4\pi}\sigma_{total}(e^+e^- \to XH) {\cal BR}(H \to
\tau^+\tau^-) \bigl( {\cal BR}(\tau \to \rho \nu_\tau) \bigr)^2 
\end{equation}
is a suitable choice.

In Fig.~\ref{rys1} the acoplanarity distribution  angle $\varphi^{*}$
of the $\rho^+ \rho^-$ decay products which was defined  in the rest
frame of the reconstructed $\rho^+ \rho^-$ pair is shown.  
A detector-like set-up is included in exactly
the same proportion as in Ref.~\cite{Desch:2003rw}.
Unobservable generator-level   $\tau^{\pm}$ rest frames are used for
the calculation of selection cuts.  The two plots represent events
selected by the differences of $\pi^\pm\pi^0$ energies, defined in
their respective $\tau^\pm$ rest frames. In the left plot, it is
required that $y_1 y_2 > 0$, whereas in the right one, events with
$y_1 y_2 < 0$ are taken. 
This figure quantifies the size of the parity
effect in an idealized condition, which we will attempt to approach
with realistic ones.  
%*******************************************************************  
\begin{figure}[!ht]
\begin{center} 
\epsfig{file=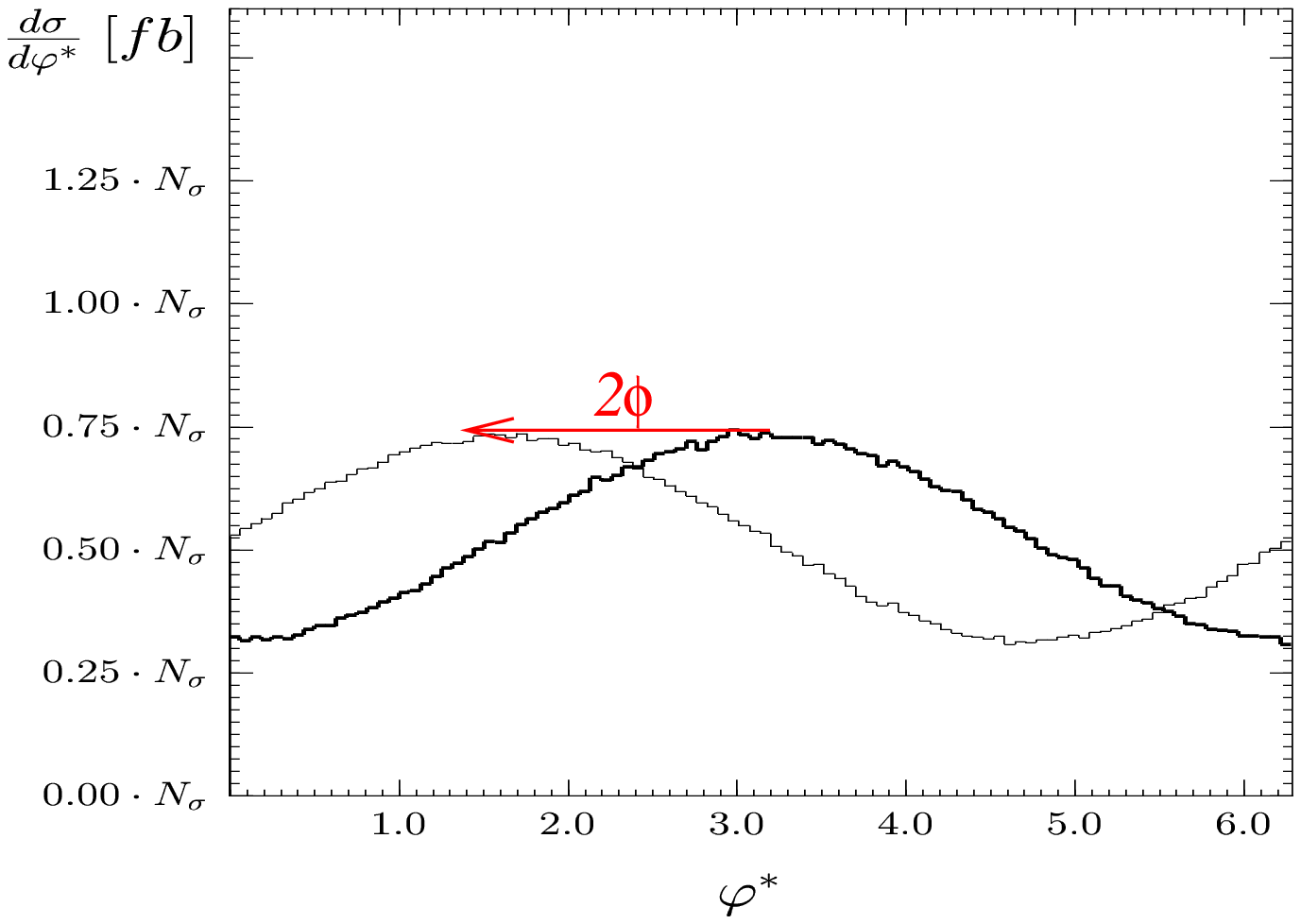,width=60mm,height=60mm}
\epsfig{file=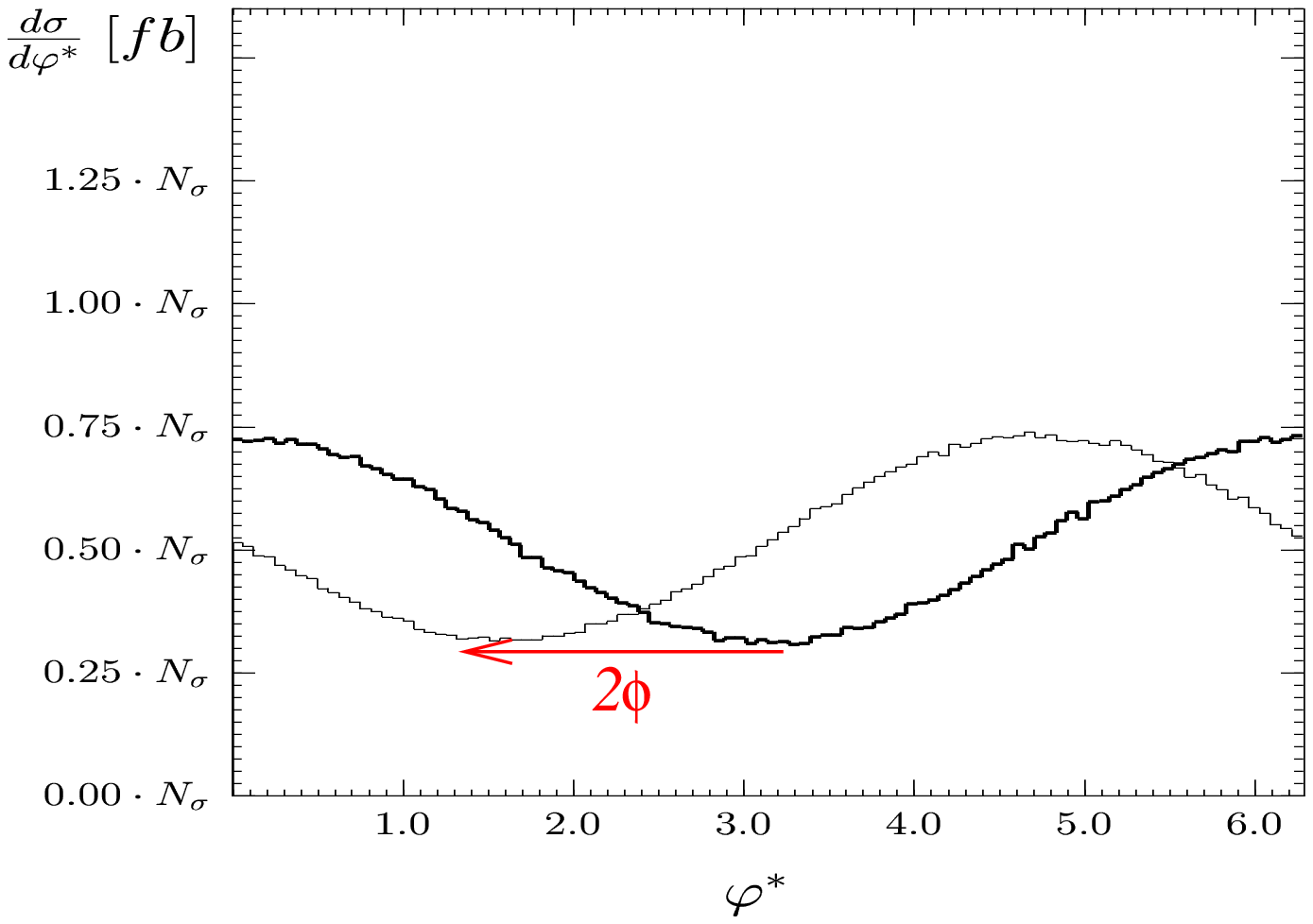,width=60mm,height=60mm}
\end{center} \vspace{-0.5 cm}
\caption  
{\it  The acoplanarity distribution  (angle $\varphi^{*}$) of the
$\rho^+ \rho^-$ decay products     in the rest frame of the $\rho^+
\rho^-$ pair. Gaussian smearing of $\pi$'s  and Higgs boson momenta,
 are included. Only events where the signs
of the  energy differences $y_1$ and $y_2$ are the same, if calculated
using the method without or with the help of the $\tau$ impact parameter
are taken.  The thick line corresponds to a
scalar Higgs boson, the thin line to a mixed one.  The left figure
contains events with $y_1 y_2 > 0$, the right one is for $y_1 y_2 <
0$. 
}
\label{rys4}  
\end{figure}  
%*******************************************************************

The size of the effect was substantially
diminished when a detector-like set-up was included for $\tau^\pm$
rest frames reconstruction as well,  see Fig.~\ref{rys4}. 
The  general shape
of the distributions however remained.
At the cost of introducing cuts, and thus reducing the number of
accepted events, we could achieve some improvement of the method, as
in Ref.~\cite{Desch:2003mw}.  If we require the signs of the
reconstructed energy differences $y_1$ and $y_2$, see Eq.~(\ref{y1y2}),
to be the same whether the method is used with or without the help of the
$\tau$  lepton impact parameter,  only  $\sim 52\%$ of events are
accepted.  The relative size of the parity effect increases. Results
are presented in Fig.~\ref{rys4}.

Both distributions clearly distinguish between decays of scalar 
or mixed Higgs boson. From the measurement of these distributions 
we can establish the  ${\cal CP}$  properties of the Higgs boson.        
More precisely, we have found, see \cite{Desch:2003rw} for details, 
that for an integrated luminosity of  $1$ $ab^{-1}$, at
$350$ $GeV$ center-of-mass energy, a high precision {\it LC} detector such as
the proposed  {\it TESLA}, should be able to measure the
scalar--pseudoscalar mixing angle for the $H \tau \tau$ coupling with
$6^{\circ}$  accuracy in the case of a Standard Model Higgs boson mass of
$120$ $GeV$.  

%%%%%%%%%%%%%%%%%%%%%%%%%%%%%%%%%%%%%%%%%%%%%%%%%%%%%%%%%%%%%%%%%%%
%                     new  section                                %
%%%%%%%%%%%%%%%%%%%%%%%%%%%%%%%%%%%%%%%%%%%%%%%%%%%%%%%%%%%%%%%%%%% 
\section*{Summary}
\label{section_5}
We have studied measurement opportunities of the ${\cal CP}$
properties of a $120$ $GeV$ Higgs boson at an $e^{+}e^{-}$
collider, \eg {\it TESLA}.
Our results show that if the Higgs sector is  ${\cal CP}$ violating then 
there is a possibility to explicitly test this ${\cal CP}$ violation trough 
spin correlations between final state particles in $H\to\tau^{+}\tau^{-}$; 
$\tau^\pm\to\rho^\pm\bar{\nu}_{\tau}(\nu_{\tau})$;
$\rho^\pm\to\pi^\pm\pi^0$ decay chains. We have found that the mixing 
scalar--pseudoscalar
angle can be determined with 
statistical precision of $6^{\circ}$ for the 
case of $e^+e^-$ collisions of $350$
$GeV$  center-of-mass energy with an integrated luminosity of $1$ $ab^{-1}$ 
and for Higgs boson mass of $120$ $GeV$.
However, we assume that the
$e^{+}e^{-}\to ZH$; 
$Z\to \mu^{+} \mu^{-}$; $H\to\tau^{+}\tau^{-}$; 
$\tau^\pm\to\rho^\pm\bar{\nu}_{\tau}(\nu_{\tau})$;
$\rho^\pm\to\pi^\pm\pi^0$ 
decay chain is background free. We
 have not introduced any cuts {\it etc.} that 
might be required to guarantee this.  

Finally, let us note that this method can be applied  to measure the
parity properties of other scalar particles, not necessarily only
Higgs boson(s).

\section*{Acknowledgments}
It is a pleasure to thank Klaus Desch, Andreas Imhof 
and  Zbigniew W\c as, with whom the work reported here was performed.
This work is partly supported by the Polish State Committee for Scientific
Research (KBN) grants Nos 5P03B09320, 2P03B00122.

\providecommand{\href}[2]{#2}

%\bibliographystyle{utphys_spires}
%\bibliographystyle{plain}
%\bibliography{TAUOLA-F}

\end{document}